# Plastic strain-induced phase transformations in silicon: drastic reduction of transformation pressures, change in transformation sequence, and particle size effect

Sorb Yesudhas[1]*, Valery I. Levitas[1,2,3]*, Feng Lin[1], K. K. Pandey[4], and Jesse Smith[5]

[1]Department of Aerospace Engineering, Iowa State University, Ames, Iowa 50011, USA

[2]Department of Mechanical Engineering, Iowa State University, Ames, Iowa 50011, USA

[3]Ames National Laboratory, Iowa State University, Ames, Iowa 50011, USA

[4]High Pressure & Synchrotron Radiation Physics Division, Bhabha Atomic Research Centre, Mumbai 400085, India

[5]HPCAT, X-ray Science Division, Argonne National Laboratory, Argonne, Illinois 60439, USA

*Corresponding authors. Email: sorbya@iastate.edu, vlevitas@iastate.edu

**Abstract:** Pressure-induced phase transformations (PTs) between numerous phases of Si, the most important electronic material, have been studied for decades. This is not the case for plastic strain-induced PTs. Here, we revealed *in-situ* various unexpected plastic strain-induced PT phenomena. Thus, for 100 nm Si, strain-induced PT Si-I→Si-II (and Si-I→Si-III) initiates at 0.4 GPa (0.6 GPa) versus 16.2 GPa (∞, since it does not occur) under hydrostatic conditions; for 30 nm Si, it is 6.1 GPa versus ∞. The predicted theoretical correlation between the direct and inverse Hall-Petch effect of the grain size on the yield strength and the minimum pressure for strain-induced PT is confirmed for the appearance of Si-II. Retaining Si-II at ambient pressure and obtaining reverse Si-II→Si-I PT are achieved, demonstrating the possibilities of manipulating different synthetic paths.

**Introduction**

Silicon is one of the most important semiconducting materials for mankind due to its extensive applications in microelectronics, integrated circuits, photovoltaics, micro, and nano-electromechanical systems (MEMS/NEMS) technologies, etc.[1,2]. Polycrystalline Si is widely used in solar panels, thin transistors, and large-scale integration manufacturing[3]. When the Si crystals or grains are scaled down to nanometer size, they show outstanding electronic, mechanical, thermoelectric, and optical properties[4-8]. Si possesses numerous high-pressure polymorphs, seven of which will be discussed here under slow compression/torsion/decompression at room temperature. In addition to the ambient semiconducting Si-I phase, the other metastable Si-III, Si-IV, and Si-XII phases are promising candidates owing to their interesting electronic and optical properties. Recently, Si-III was synthesized from Si-I at 14 GPa and ~900 K and quenching over 3 days or through a chemical pathway in a Na-Si mixture at 9.5 GPa and ~1000[9]. Similar studies reveal that Si-III is an ultranarrow band gap semiconductor (with a band gap $E_g$~30 eV) with much lower thermal conductivity compared with Si-I and is found to have the potential for infrared plasmonic applications[10]. High-pressure torsion (i.e., large plastic shear under pressure) processing of Si at 24 GPa and 10 anvil rotations were used to obtain nanostructured metastable Si-III and Si-XII phases[11,12]. These are too high pressures for industrial applications; we will show that we can drastically reduce them by plastic straining of nano-grained Si-I.

The *in-situ* hydrostatic[13-17], uniaxial[7,18-20], and indentation[21,22] studies of Si, including nanoindentation, nanospheres, and micropillars, under compression-decompression, are very extensive. However, many processes, like friction, machining, dicing, lapping, scribing, and polishing of Si wafers and polycrystals, are accompanied by large plastic shear, which may strongly alter the PT pressures[23-28]. As formulated in[23], there is a fundamental difference between PTs under hydrostatic or non-hydrostatic compression below the yield (pressure- and stress-induced PTs) and PTs during plastic flow (strain-induced PTs). The pressure-induced PTs are initiated by nucleation at the pre-existing defects, like dislocations or grain boundaries, whereas the plastic strain-induced PTs occur at new defects constantly generated during the plastic flow. Plastic strain-induced PTs may occur under much lower pressure, follow strain-controlled (instead of time-controlled) kinetics, and provide new avenues to explore hidden phases, which cannot be obtained under hydrostatic compression[23-28]. The plastic strain-induced PTs require completely different experimental characterization and thermodynamic and kinetic descriptions.

While there are hundreds of publications on PTs in Si under quasi-hydrostatic conditions in a diamond anvil cell DAC (DAC) and nanoindentation, only one quite old article on PTs under compression and shear in rotational DAC (RDAC)[24,29], utilizing only optical and electric resistivity measurements, without *in-situ* XRD, wherein the authors claim a PT sequence, Si-I→III→II. Later[30], using a TEM study, it was claimed that Si-I→IV→III occurs at 2-4 GPa and Si-I→III→II at higher pressure. These claims contradict our *in-situ* synchrotron x-ray diffraction (XRD) results in Fig. 1. Also, it is known that particle/grain size reduction increases the PT pressure for pressure-induced PTs[31]; however, there are no similar studies for strain-induced PTs.

Here, we present *in-situ* synchrotron XRD studies of various plastic strain-induced PTs for Si with three different Si particle sizes, ~1 µm, 100 nm, and 30 nm, compared with pressure-induced PTs and reveal several unexpected phenomena and transformation paths. Various possible fundamental and technological applications are analyzed.

**Materials and Experimental Methods**

Hydrostatic (with pressure transmitting medium (PMT)) and plastic (without PMT) compressions in DAC and torsion in the RDAC were conducted on Si with particle/grain sizes of d = 1 µm, 100 nm, and 30 nm. Single crystal silicon purchased from Sigma-Aldrich was powdered to 1 µm size fine powders by grounding with a pestle and mortar. The Si powder with < 100 nm particle size (TEM) and ≥ 98 % pure sample was purchased from Sigma-Aldrich (CAS No.:7440-21-3). Also, Si powder with 30 nm particle size was purchased from Meliorum Technologies, Inc. The high-pressure experiments were carried out using symmetric-type DAC and RDAC with 300 µm and 400 µm culet diameters. We have used both polished and rough diamonds [32] for this study, and the description of the RDAC can be found in[33].

Both stainless steel (S.S.) and Cu gaskets with an initial thickness of 70-150 µm were utilized with the hole (sample) diameters of 120-220 µm. The non-hydrostatic sample loading was carried out with Cu and S.S. gaskets; no PTM was used. All the high-pressure X-ray diffraction experiments were carried out at the 16-ID-B beam line, HPCAT utilizing Advanced Photon Source with X-ray wavelengths 0.4133 Å and 0.3445 Å. The X-ray beam with spot size 5µm x 4µm was scanned along the diameter of the sample with step sizes 8 µm and 10 µm at various pressures. Pressure at each sample point in each phase was determined using equations of the state of Si phases obtained using He PTM [34] with a ruby pressure scale[35]. The XRD patterns were refined using GSAS-II software to extract the lattice parameters, volume fractions of phases, and texture by the Rietveld refinement method[36]. The 2D XRD image was converted to a 1D pattern using dioptase software[37]. The spherical harmonics texture model was used to fit the high-pressure Si phases. The radial thickness profile of the sample was determined using the X-ray absorption method.

**Results and discussions**

***Theoretical predictions.*** It is justified in our analytical model[23] and phase field simulations[38,39] that plastic strain-induced PT occurs by nucleation at the tip of a dislocation pileup as the strongest possible stress concentrator. According to the classical Eshelby model[40], all components of stress tensor ($\sigma$) at the tip of edge dislocation pileup, modeled as a superdislocation, are $\sigma \sim \tau l$, where $\tau$ is the applied shear stress limited by the yield strength in shear $\tau_y$, $l$ is the length of the dislocation pileup. Since $l$ is traditionally limited by the fraction of the grain size $d$ (e.g., $l=0.5d$), the main conclusion in[23] was that the greater the grain size, the stronger the stress concentration and consequently, the reduction in PT pressure. This is opposite to what we found in experiments, e.g., for α-ω PT in Zr[28]. To eliminate this contradiction, we will utilize our phase field[38,39], molecular dynamics[41], and concurrent atomistic-continuum simulations[42]. In contrast to the analytical solution utilized in[23], $l$ is not related to the grain size since most dislocations are localized at the grain boundary producing a step (superdislocation) (Fig. S1), with effective length $l=Nb<<d$, where $N$ is the number of dislocations in a pileup and $b$ is the magnitude of the Burgers vector. Then we can use the stress field of a superdislocation, i.e., $\sigma \sim \mu Nb/r$, where $\mu$ is the shear modulus, and $r$ is the distance from the midpoint of a pileup. The number of dislocations in a pileup in the first approximation, $N \sim \tau = \tau_y$, i.e., $\sigma \sim \mu \tau_y b/r$. At the same time, $\tau_y$ increases with the decrease in $d$ according to the Hall-Petch relationship, $\tau_y = \tau_0 + kd^{-0.5}$, where $\tau_0$ and $k$ are material parameters. Thus, $\sigma \sim \mu (\tau_0 + kd^{-0.5})b/r$. That is why the stress concentration increases and the minimum pressure for the strain-induced PT $p_\varepsilon^d$ decreases with decreasing grain size. This prediction is counterintuitive because, under hydrostatic conditions, the PT pressure increases with the grain size reduction[31], which we also confirm below. However, for very small grain sizes, the above reasoning is not valid because the yield strength decreases with a further grain size reduction (the inverse Hall-Petch effect), and grain boundary sliding competes with the dislocation pileup formation. Then based on equation $\sigma \sim \mu Nb/r$ and decreasing $N$ with decreasing grain size, we conclude that the stress concentration decreases and minimum pressure for the strain-induced PT $p_\varepsilon^d$ increases with decreasing grain size in the range of validity of the inverse Hall-Petch effect. These predictions, including crossover in PT pressure with decreasing grain size, will be confirmed below experimentally, along with many other results.

***Experimental results***

A summary of observed experimental results for hydrostatic, plastic compression in DAC, and torsion in RDAC for three types of Si particles are presented in Figs. 1a-b, and the results are discussed below.

(a) Hydrostatic Compression
Micron Si: I $\xrightarrow{13.5\ GPa}$ I+II+XI $\xrightarrow{15.3\ GPa}$ XI $\xrightarrow{16.5\ GPa}$ V $\xrightarrow{0\ GPa}$ XII+III     (PTM: He)
100 nm Si: I $\xrightarrow{16.2\ GPa}$ I+II+XI $\xrightarrow{18.1\ GPa}$ XI $\xrightarrow{19.3\ GPa}$ V $\xrightarrow{0\ GPa}$ XII+III     (PTM: He)
30nm Si:   I $\xrightarrow{14.6\ GPa}$ I+XI $\xrightarrow{19.8\ GPa}$ I+XI+V $\xrightarrow{21.6\ GPa}$ I+V $\xrightarrow{23.2\ GPa}$ V $\xrightarrow{0\ GPa}$ a     (PTM:He)

Non-hydrostatic Compression
Micron Si: I $\xrightarrow{2.5\ GPa}$ I+II $\xrightarrow{8.8\ GPa}$ I+II+XI $\xrightarrow{14.1\ GPa}$ XI $\xrightarrow{14.2\ GPa}$ XI+V $\xrightarrow{14.8\ GPa}$ V $\xrightarrow{0\ GPa}$ XII+III     (S.S.)
Micron Si: I $\xrightarrow{2.2\ GPa}$ II     (Cu)
100 nm Si: I $\xrightarrow{0.4\ GPa}$ I+II $\xrightarrow{13\ GPa}$ I+II+XI $\xrightarrow{14.4\ GPa}$ II+XI $\xrightarrow{17.2\ GPa}$ XI+V $\xrightarrow{18.2\ GPa}$ V $\xrightarrow{0\ GPa}$ III     (Cu)
30nm Si:   I $\xrightarrow{6.1\ GPa}$ I+II $\xrightarrow{9.9\ GPa}$ I+II+XI $\xrightarrow{11.8\ GPa}$ I+II+XI+V $\xrightarrow{12.6\ GPa}$ I+XI+V
         II &I+II $\xleftarrow{0\ GPa}$ V $\xleftarrow{13.4\ GPa}$     (Cu)

Torsion
100 nm Si: I $\xrightarrow{5°,0.4\ GPa}$ I+II $\xrightarrow{5°,0.6\ GPa}$ I+II+III $\xrightarrow{82°,8.3\ GPa}$ I+II+III+XI $\xrightarrow{12\ GPa}$ II+XI $\xrightarrow{0\ GPa}$ XII+III (Cu)
100 nm Si: I $\xrightarrow{2.2\ GPa}$ I+II $\xrightarrow{31.8°,4.2\ GPa}$ I+II+III $\xrightarrow{0\ GPa}$ I+II&I+III&I+II+III   Rough anvils     (S.S.)
100 nm Si: I $\xrightarrow{9.1°,3.6\ GPa}$ I+II $\xrightarrow{0\ GPa}$ I   Rough anvils     (S.S.)

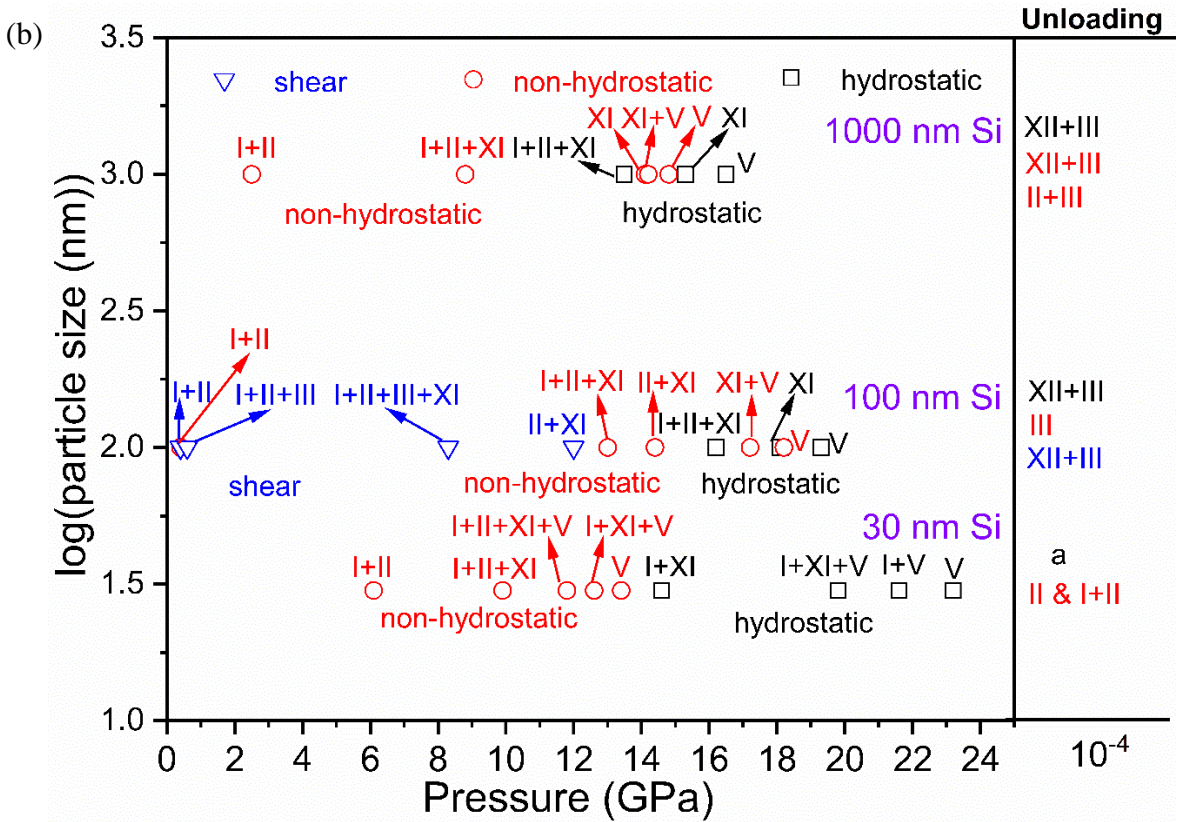

Fig. 1. Schematic of a sequence of PTs in Si with different particle sizes.

**(a)** Sequence of PTs in Si with different particle sizes (1000 nm, 100 nm, and 30 nm) under hydrostatic and non-hydrostatic (plastic) compressions and torsion inside RDAC. The 0 GPa represents the unloading of Si to $10^{-4}$ GPa. I, II, III, XI, V, XII, and a represent the diamond cubic (S.G: Fd$\bar{3}$m), tetragonal (I4$_1$/amd), bcc (Ia$\bar{3}$), orthorhombic (Imma), simple hexagonal (P6/mmm), rhombohedral (R$\bar{3}$), and amorphous, respectively. Different results are separated by & are for different regions of the same sample. The pressure transmitting medium (PTM) and gasket types used in the experiment are shown at the end of the figure.

**(b)** Relationship between the PT pressure and particle size of Si under hydrostatic (black squares) and non-hydrostatic (red circles) compressions and torsion inside RDAC (blue triangles). Different phases observed after unloading are shown on the right side of the figure.

*Hydrostatic compression*

Under *hydrostatic compression*, a decrease in particle size, $d$, from micron to 100 nm increases transition pressures for all PTs by ~3 GPa, which is related to the reduction in the probability of having a strong stress concentrator due to pre-existing different dislocation configurations with reducing $d$ (Fig. 1). The same PT sequence is observed in the literature for micron-size Si powder particles[14]. The Si-I phase is stable up to 23.2 GPa and then transforms directly to Si-XI and Si-V by skipping the intermediate Si-II phase. During pressure release, a-Si appears instead of Si-XII and Si-III (Fig. 1). Our findings for $d$ = 30 nm particles agree with the results in[31], but the results described below are observed for the first time. Thus, the transition pressure of Si-XI reduces by ~1.6 GPa, but Si-V increases by ~0.8 GPa in comparison with 100 nm Si.

Note that perfect Si-I crystal shows transverse acoustic phonon instability at 18.3 GPa, according to the first-principle simulations[43], i.e., it cannot exist above this pressure. The above results can be rationalized by the increasing role of the surface energy for 30 nm particles. Thus, the surface energy strongly increases during Si-I→Si-II PT, suppressing this PT for all particles. This effect increases with a reduction in particle size, and for 30 nm Si, Si-XI and V appear before Si-II due to their smaller surface energy. These results raise theoretical challenges for their quantitative descriptions.

*Plastic strain-induced PTs*

Numerous breakthrough results have been obtained in Figs. 1-4:

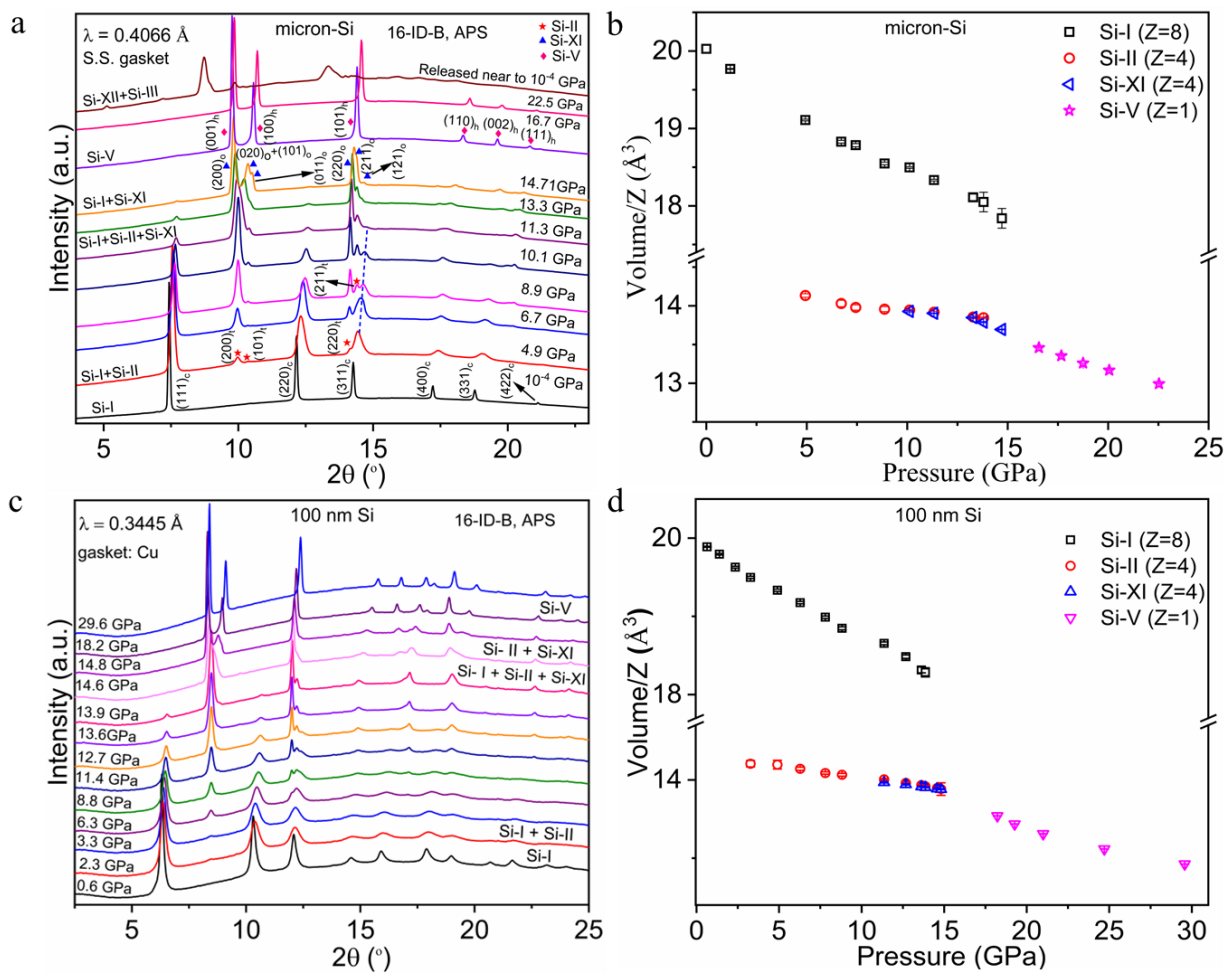

**Fig. 2. Plastic strain-induced PTs in the micron (a,b) and 100 nm (c,d) Si under non-hydrostatic compression with smooth diamonds.** (**a**) XRD of micron-size silicon at various pressures under non-hydrostatic compression at 8 µm away from the sample center. The Bragg planes for the Si-II, XI, and V phases are identified by star, triangle, and diamond symbols, respectively. The evolution of the (311) plane of the Si-I is shown as a dashed line. The phase coexistence at various pressures is also shown. The Si-II is observed at 4.9 GPa at ~10 µm from the sample center and at 2.5 GPa close to the gasket. (**b**) Volume/Z vs. pressure of micron-Si. (**c**) XRD of 100 nm Si at various pressures at the sample center. The phase coexistence at various pressures is shown. (**d**) Volume/Z *vs.* pressure of 100 nm Si. The pressure was calculated by the equation of state of Si phases of 100 nm Si and 1 µm Si obtained by He PTM with ruby calibrant. The pressure of the Si-II phase of 1 µm Si was calculated using the Si-II EOS in[14].

(a) For 100 nm Si, plastic straining reduces Si-I→Si-II PT pressure from 16.2 GPa under hydrostatic loading to $p_\varepsilon^d = 0.4$ GPa, i.e., by a factor of 40.5! (Figs. 1 and 4). This is also 26.3 times below the phase equilibrium pressure of 10.5 GPa[13]. For micron-Si, the pressure reduction for this PT is from 13.5 to 2.5 GPa, by a factor of 5.4; for 30 nm particles, it is from ∞ (since Si-II does not appear) to 6.1 GPa (Figs. 1-3). The effect of plastic strain is much weaker for the appearance of Si-XI: while for micron and 100 nm particles, Si-XI appears simultaneously with Si-II under hydrostatic loading, $p_\varepsilon^d$ for Si-XI is by 7.0 and 8.5 GPa larger than for Si-II (Figs. 1,2, and 4). For 30 nm Si, PT pressure for the appearance of Si-XI reduces from 14.6 to 9.9 GPa (Figs. 1,3b,c). The effect of plastic strain on the appearance of Si-V is weaker than for Si-II and Si-XI (Figs. 1-4). A highly complex phase coexistence with texture is observed in the plastically compressed samples (Fig. 1, 2b,d, 3b, and S2).

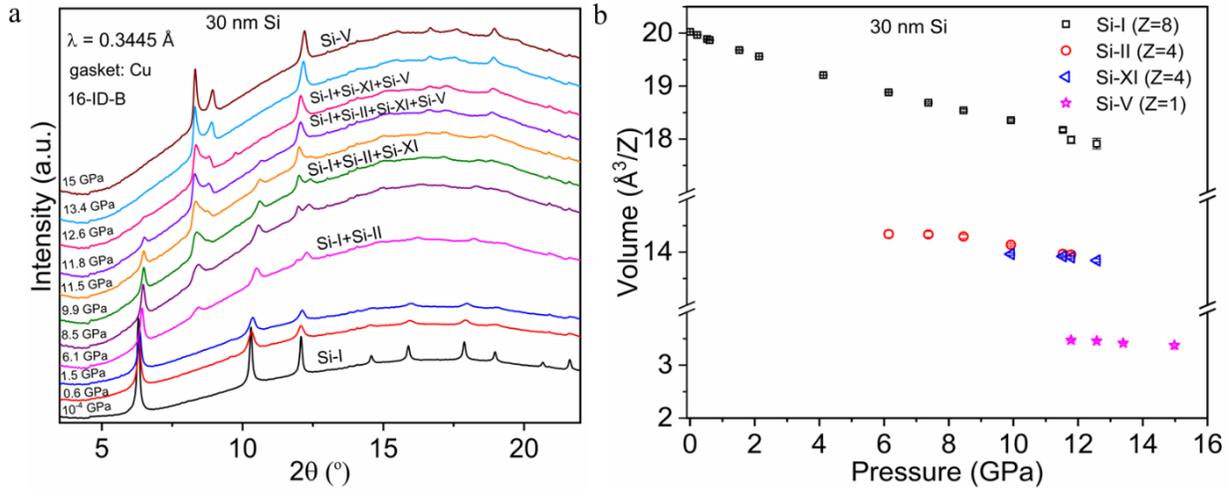

**Fig. 3. Plastic strain-induced PTs under non-hydrostatic compression in 30 nm Si at the sample center.** **(a)** High-pressure XRD pattern of 30 nm Si at various pressures with smooth diamonds and Cu gasket. The phase coexistence at various pressures is shown. **(b)** Volume/Z *vs.* pressure for 30 nm Si. The pressure was calculated by the equation of state of Si phases of 30 nm Si obtained by He PTM with ruby calibrant.

(b) Our theoretical predictions on grain size dependence of $p_\varepsilon^d$ is confirmed for the appearance of Si-II, i.e., there is a reduction in $p_\varepsilon^d$ from micron to 100 nm particles and then increase for 30 nm particles (Figs. 1-4). For the next PTs, to Si-XI and V, this is not expected because the actual grain size after previous PT(s) is unknown and may not be related to the initial particle size; also, PT occurs in a mixture of phases instead of single phase, and phase interfaces may serve as additional obstacles for dislocation pileup. Determination of grain size evolution that we plan in future work will allow us to shed light on this problem.

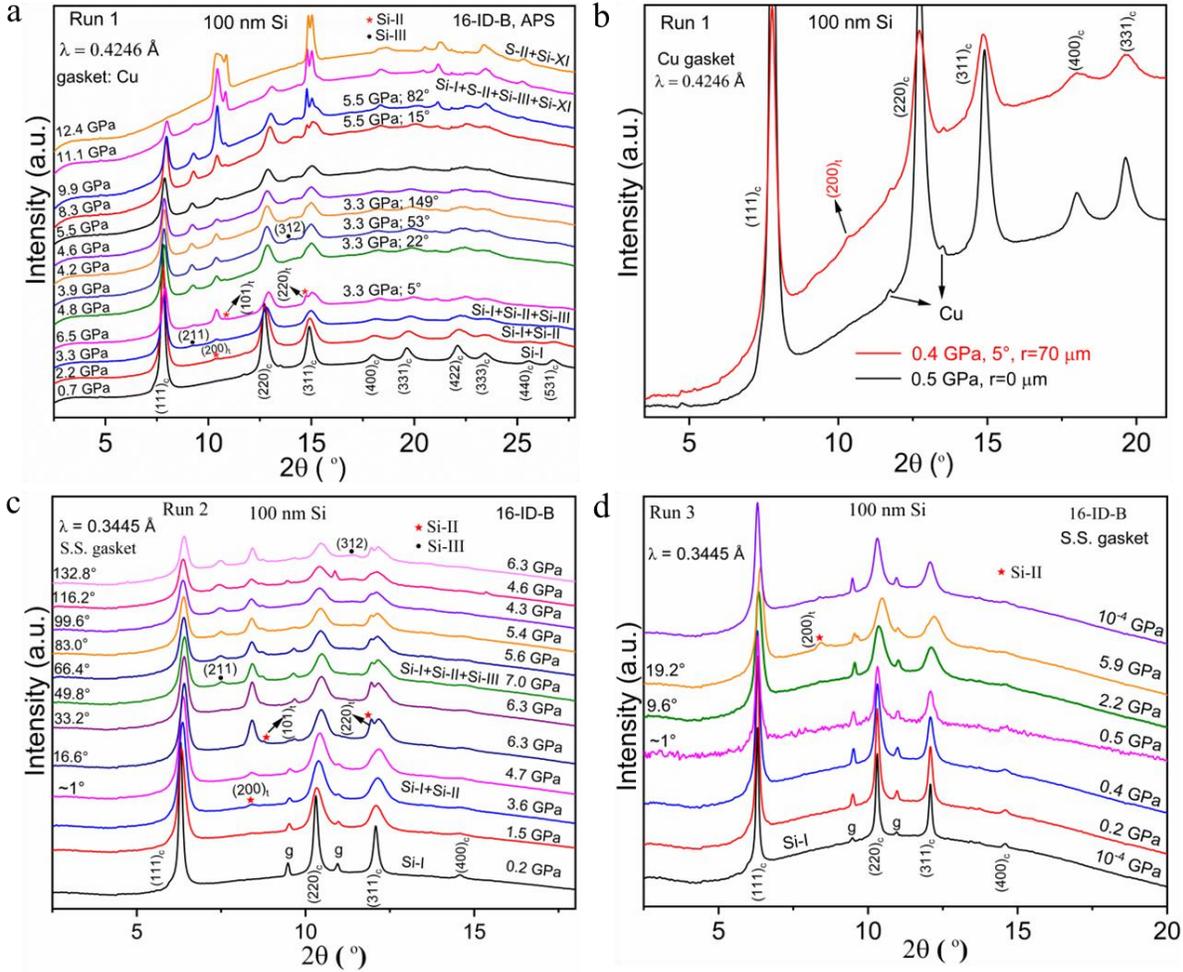

**Fig. 4: Plastic strain-induced PTs in 100 nm Si with torsion inside the RDAC.** (**a**) XRD patterns of 100 nm Si within Cu gasket at various pressures and anvil rotations with smooth diamonds at the sample center. The suffix 't' of the Bragg planes denotes the tetragonal Si-II. The torsion is applied at 3.3 GPa and 8.3 GPa. The designation, e.g., 3.3 GPa; 22° means rotation angle starting from the pressure at the center of 3.3 GPa; actual pressure is shown at the left side of the plot. The phase coexistence at various pressures is shown. (**b**) The initiation of Si-I→Si-II PT at 0.4 GPa; g is the gasket peak. r =0 and r = 70 µm denote the XRD patterns collected at the sample center and 70 µm away from the center, respectively. (**c**) XRD patterns of 100 nm Si at various

pressures and torsions with rough diamonds and the S.S. gasket. Si-II was initiated at 3.6 GPa at the gasket center. The piston side diamond anvil was rotated continuously up to 132.8° at 1 hour per rotation. **(d)** XRD patterns of 100 nm Si at selected pressures and torsions with rough diamonds at the sample center with S.S. gasket. The sample was pressurized up to 0.5 GPa, and the shear was applied by continuously rotating one of the anvils up to 19.2° at 1 hour per rotation. The Si-II was initiated at 0.5 GPa at the gasket center. The pressure was released to the ambient from 5.9 GPa. The Si-I phase is recovered after the pressure release, i.e., the reverse Si-II→Si-I PT was observed. The pressure is calculated using the equation of the state of the Si-I phase of 100 nm obtained by He PTM with ruby calibrant.

(c) For PT I→II, $p_\varepsilon^d = 0.4$ GPa in 100 nm Si is the same for compression in DAC and torsion in RDAC (Figs. 1-4). Since the plastic strain tensor and its path are quite different for compression and torsion, which means that $p_\varepsilon^d$ is independent of the plastic strain tensor and its path. A similar result we obtained previously for α-ω PT in Zr[28,32], i.e., this may be a general rule for different classes of materials. This rule reveals that the physics of strain-induced PTs under compression in DAC and torsion in RDAC do not differ fundamentally, and strain-induced PTs can also be studied in DAC. Of course, RDAC has a strong advantage in allowing various controllable pressure-shear loading programs, particularly at constant pressure, close to $p_\varepsilon^d$, whereas in DAC, pressure significantly grows during compression. This is crucial for not only initiating but also completing the desired PTs at low pressure.

(d) Under the plastic shear of 100 nm Si in RDAC, Si-III initiates at 0.6 GPa, while it does not appear under hydrostatic and plastic compression (Figs. 1 and 4b).

(e) For 100 nm Si, under torsion, PT occurs in the sequence Si-I→I+II→I+II+III (Figs. 4a-d), in contrast to different PT sequences suggested in[24,29,30] not supported by the *in-situ* XRD study. The direct transition of Si-I→Si-III requires breaking 40% of covalent bonds of the cubic diamond lattice of Si-I, which can be furnished by shear strain[24]. However, whether Si-III is produced from Si-I or II needs to be clarified.

(f) The coexistence of Si-I, Si-II, Si-XI, and Si-V under compression of 30 nm Si and Si-I, Si-II, Si-III, and Si-XI under torsion of 100 nm Si were observed for the first time.

One of the longstanding puzzles in Si is stabilizing the high-pressure Si-II phase to the ambient for *in-situ* studies and potential engineering applications. In DAC experiments, Si-II was not retained at ambient conditions, even though the first-principle simulations[44,45] show the metastability of Si-II. Based on obtained experimental results and multiscale simulations[27], we were able to design a low-pressure compression-torsion loading path for 100 nm Si utilizing stainless steel and Cu gaskets and rough diamonds; we could retain a small amount of Si-II at normal pressure in several regions (Figs. 4 and S3). Interestingly, in our plastic compression experiment on 30 nm Si, we obtained Si-V→Si-II phase near ambient, and after the complete pressure release, the Si-II phase with partial Si-I phase was observed (Fig. S4). A significant amount of Si-II was retained during Si-V→Si-II unloading of micron Si after plastic compression (Fig. S5). Retaining a large amount of Si-II will allow us to study it with traditional *ex-situ* methods (SEM, TEM, mechanical properties, etc.) and eventually discover its potential applications, as it happened with Si-III[10]. We have also observed reverse Si-II→Si-I PT, which was never seen before since Si-II transformed to Si-III and XII during unloading. Si-I+Si-III and Si-I+Si-II+Si-III phase coexistences at several regions under compression and shear in RDAC were also observed (Fig. 4a-d). Astonishingly, we recovered nano-Si-III when unloading Si-V of 100 nm Si subjected to plastic compression in DAC (Fig. S6), which was never done before. There are various potential applications of the obtained results.

(a) Drastic reduction in PT pressures due to plastic deformation and its nontrivial dependence on the particle size, as well as change in transformation paths open (i) new basic direction in the multiscale theoretical description of these phenomena, e.g., by further developing approaches reviewed in,[31] and (ii) applied direction in developing the scientific foundation for new plastic strain- and defect-induced synthesis and retrieving the desired nanostructured pure phases or mixture of phases (nanocomposites) with optimal electronic, optical, and mechanical properties. This can be done at low pressure, room temperature, in a short time, and without a catalyst. Due to multiple PTs, Si represents an ideal model material for developing similar approaches for other strong and brittle semiconductors (Ge, $Si_xGe_{1-x}$, GaAs, InSb, GaSb, etc.), graphite-diamonds (cubic, hexagonal, orthorhombic, etc.) and similar BN systems.

(b) In particular, while obtaining pure Si-III after quenching from 14 GPa and 900 K over three days or in Na-Si system at 9.5 GPa and 1000 K[16] or by 2 compression/slow (within 4 hours) decompression cycles to 13 GPa and room temperature,[17] it was mentioned in[16] as a promising result that Si-III appeared at 7 GPa and 1000 K. We obtained Si-III at 0.6 GPa, room temperature, and anvil rotation of 5° during minutes, and various synthetic compression-torsion-unloading paths can be developed to obtain pure Si-III at ambient pressure. Since the transformation rate is proportional to the strain rate[1,31], it can be increased by increasing the rotation rate.

(c) Obtained results can also be used for quantitative modeling and optimization of surface processing (polishing, turning, scratching, etc.) of strong brittle semiconductors and developing regimes of ductile machining by utilizing PT to ductile (Si-II and a-Si) phases[23,25]. Current molecular dynamics modeling on cutting polycrystalline Si-I show the formation of Si-II and a-Si at ~10 GPa under shear, i.e., close to the values for a single crystal under hydrostatic conditions.[37] The main reason is that atomistic simulations are limited to very small grain sizes, which are in the range of the validity of the inverse Hall-Petch effect, like for 30 nm Si in Fig. 1. For 100 nm particles, this pressure can be drastically reduced to sub-GPa (Fig. 1).

(d) Since the PTs in Si were typically obtained at >10 GPa, they were neglected in typical normal and relatively low-pressure applications. Observed very low PT pressures warn that some PTs may occur, e.g., in NEMS/MEMS, during contact interactions and friction, and should be taken into account or/and avoided.

(e) Our results, especially with rough diamonds, can be scaled up to a larger volume using high-pressure torsion with rotating metallic/ceramic anvils[22].

(f) RDAC studies can also be used for *in-situ* studying processes occurring during high-pressure torsion. Traditional high-pressure torsion cannot be monitored *in-situ*; it is studied *ex-situ* after completing the entire or part of the process and unloading. Since Si phases after unloading differ from during loading (Fig. 1), the final product does not characterize any PT during the loading. Even for other material systems without PTs at pressure release, the *in-situ* study is much more representative, better characterized, and precise because it allows getting much more experimental points at different loadings, characterize parameters (e.g., pressure) in each point (instead of force divided by total area), and avoids damage of the brittle materials at normal conditions.

(g) Large transformation strain $\varepsilon_t = \{-0.514; 0.243; 0.243\}$ with volumetric strain of -0.249 for Si-I→Si-II PT causes large transformation-induced plasticity (TRIP) and energy absorption. They can be utilized in Si nano- and micro-inclusions to obtain transformation toughening in strong brittle materials, like armor ceramics. Particle size and type of loading can control the PT pressure. TRIP may be very significant in the shear-PT bands.[38]

**Acknowledgments:** Support from NSF (CMMI-1943710) and Iowa State University (Vance Coffman Faculty Chair Professorship) is greatly appreciated. This work is performed at HPCAT (Sector 16), Advanced Photon Source (APS), and Argonne National Laboratory. HPCAT operations are supported by DOE-NNSA's Office of Experimental Science. The Advanced Photon Source is a U.S. Department of Energy (DOE) Office of Science User Facility operated for the DOE Office of Science by Argonne National Laboratory under Contract No. DE-AC02-06CH11357.

**Author contributions:** SY and FL performed experiments. SY collected and analyzed the data. VIL conceived the study, supervised the project, and developed theoretical models. KKP and JS assisted with experiments. VIL and SY prepared the initial manuscript. All authors contributed to discussions of the data and the manuscript.

**Competing interests**

The authors declare no competing interests.

**Data availability**

The data that support the findings of this study are available from the corresponding authors upon request.

# Supplementary Material

**Plastic strain-induced phase transformations in silicon: drastic reduction of transformation pressures, change in transformation sequence, and particle size effect**

Sorb Yesudhas[1*], Valery I. Levitas[1,2,3*], Feng Lin[1], K. K. Pandey[4], and Jesse Smith[5]

[1]Department of Aerospace Engineering, Iowa State University, Ames, Iowa 50011, USA

[2]Department of Mechanical Engineering, Iowa State University, Ames, Iowa 50011, USA

[3]Ames National Laboratory, Iowa State University, Ames, Iowa 50011, USA

[4]High Pressure & Synchrotron Radiation Physics Division, Bhabha Atomic Research Centre, Mumbai 400085, India

[5]HPCAT, X-ray Science Division, Argonne National Laboratory, Argonne, Illinois 60439, USA

*Corresponding authors. Email: sorbya@iastate.edu, vlevitas@iastate.edu

**Contents**

1. **Supplementary Discussion**
2. **Supplementary Figures**
3. **References**

**Supplementary Discussion**

(i) Phases and phase transformations in silicon under hydrostatic loading and unloading.
(ii) A new sequence of phase transformation in 100 nm Si.

**Figure captions**

Fig. S1. Dislocation pileups produce a step at the grain boundary or phase interface that causes a phase transformation.
Fig. S2. Rietveld refinement of non-hydrostatic compression of 30 nm Si at 11.8 GPa at r = 0 µm.
Fig. S3: XRD patterns of pressure released sample of 100 nm Si after compression and shear inside RDAC.

Fig. S4. XRD patterns of sample of 30 nm Si after non-hydrostatic compression and complete pressure release.
Fig. S5. Rietveld refinement of pressure released sample of micron-Si.
Fig. S6. Rietveld refinement of pressure released 100 nm Si sample after plastic compression.

**Supplementary Discussion**

(i) **Phases and phase transformations in silicon under hydrostatic loading and unloading**

The following phases of Si are obtained below 80 GPa.

**Loading**[1]**:** Si-I, cubic diamond, pressure range: 0-13 GPa, S.G: $Fd\bar{3}m$, semiconductor;
Si-II, β-tin, pressure range: 13.1-14.3 GPa, S.G: $I4_1/amd$, metal;
Si-XI, orthorhombic, pressure range: 13.1-16 GPa, S.G: Imma, metal;
Si-V, simple hexagonal (sh), pressure range: 15.8-40.9 GPa, S.G: $P6/mmm$;
Si-VI, orthorhombic, pressure range: 42.1-46 GPa, S.G: Cmca;
Si-VII, hexagonal, pressure range: 42.1-94.4 GPa, S.G: $P6_3/mmc$.

**Unloading:**
**Amorphous phases**[2]: Low- and high-density a-Si.
**Rapid decompression**[3]: Si-VIII, tetragonal, S.G: $P4_12_12$ and Si-IX tetragonal, S.G: $P4_222$.
**Slow decompression**[4-6]: Si-XII, rhombohedral (r8), S.G: $R\bar{3}$, semiconductor, and Si-III, bcc (BC8), S.G: $Ia\bar{3}$, ultra-narrow band gap semiconductor.
**Heating Si-III**[5]: Hexagonal diamond (hd), S.G: $P6_3mc$, metallic.

7 of them were observed in our study (Fig. 1).

(ii) **A new sequence of phase transformations in 100 nm Si**

The plastic strain-induced PT studies have been carried out on 100 nm Si with Cu and S.S. gaskets of different thicknesses. The XRD patterns of strain-induced PTs study using a Cu gasket with 150-micron thickness inside RDAC is shown in Figs. 4a-b. The sample was compactly packed inside a Cu gasket pre-indented to 150 µm thickness using 400 µm diameter smooth diamonds, and a 250 µm hole diameter was made using an HPCAT, APS laser drilling machine. The XRD spectra were collected along the culet diameter with a 10-micron step size. The sample pressure is maximum at the sample center. The appearance of a new peak at $2\theta = 10.36°$ confirms the initiation of the Si-II phase at the sample center. The Si-I→Si-II phase transformation is initiated at 2.2 GPa at the sample center (Fig. 4a). The initiation pressure of Si-II is much smaller away from the sample center due to the large plastic strain. The initiation of Si-II and Si-III phases are at 0.4 GPa and 0.6 GPa, respectively, away from the sample center. The Si-I and Si-II phases coexist at 2.2 GPa. At 3.3 GPa, another peak appeared close to the (111) peak of the Si-I phase. The Si-II and Si-III phases evolve with pressure; correspondingly, the phase fraction of the Si-I phase decreases. The Si-I, Si-II, and Si-III phases coexist at 3.3 GPa. Shear is applied at 3.3 GPa by the anvil rotation to various degrees at one rotation/hour. The phase fractions of Si-II and Si-III phases increase with pressure. The sample pressure is calculated using the equation of the state of the Si-I and Si-II phases. The calculated pressure at an initial pressure of 3.3 GPa and 149° is 4.6 GPa. The sample is pressurized to 5.5 GPa, and the shear is applied again. The Si-XI phase is observed at 5.5 GPa and 82°, and all four phases coexist, viz., Si-I, Si-II, Si-III, and Si-XI. With further compression, at 12.4 GPa, Si-I and Si-III phases disappeared, and Si-II and Si-XI coexist. The Si-III phase intermediate between Si-II and Si-XI is observed for the first time during loading. At some regions away from the sample center, Si-II and Si-III phases were initiated simultaneously. Also, in one or two regions Si-III phase is initiated before the Si-II. The appearance of the Si-III phase is reproduced in all our shear experiments with Cu and S.S. gaskets.

**Supplementary Figures**

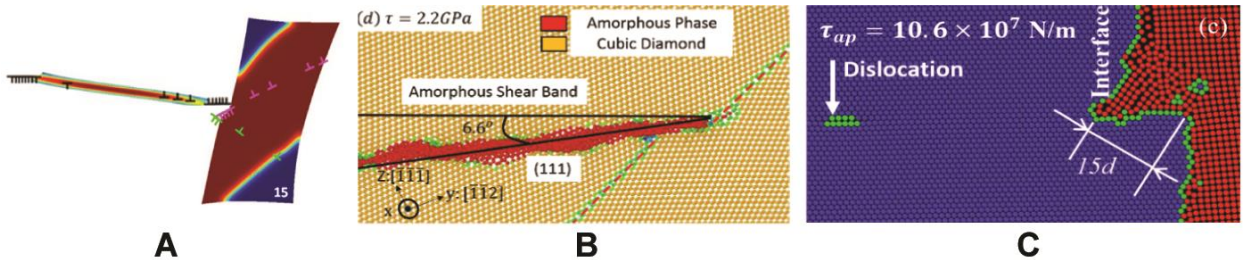

**Fig. S1. Dislocation pileups produce a step at the grain boundary or phase interface that causes a phase transformation.** (**A**) Dislocation pileup in the left grain produces a step at the grain boundary and cubic to tetragonal PT and dislocation slip in the right grain. Phase-field approach results in from[7]. (**B**) Dislocation pileup in the right grain produces a step at the grain boundary in Si-I and amorphization in the left grain. Molecular dynamics results in from[8]. (**C**) Step at the phase interface boundary consisting of 15 dislocations, causing cubic to hexagonal PT. The atomistic portion of the concurrent continuum-atomistic approach from[9]. Adopted with changes from[7-9] with permissions.

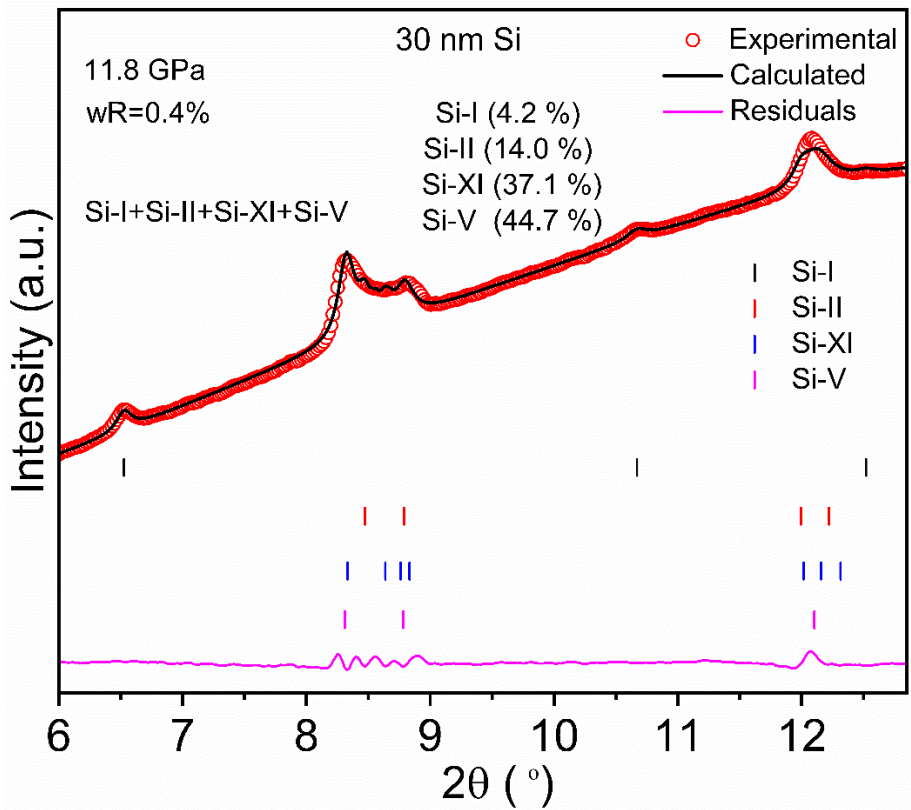

**Fig. S2. Rietveld refinement of non-hydrostatic compression of 30 nm Si at 11.8 GPa at r = 0 µm.** Rietveld refinement of non-hydrostatic compression of 30 nm Si at 11.8 GPa at the sample center, r = 0 µm (the sample center) compressed inside a Cu gasket. The phase fractions (in %) of Si-I, Si-II, Si-XI, and Si-V phases are shown.

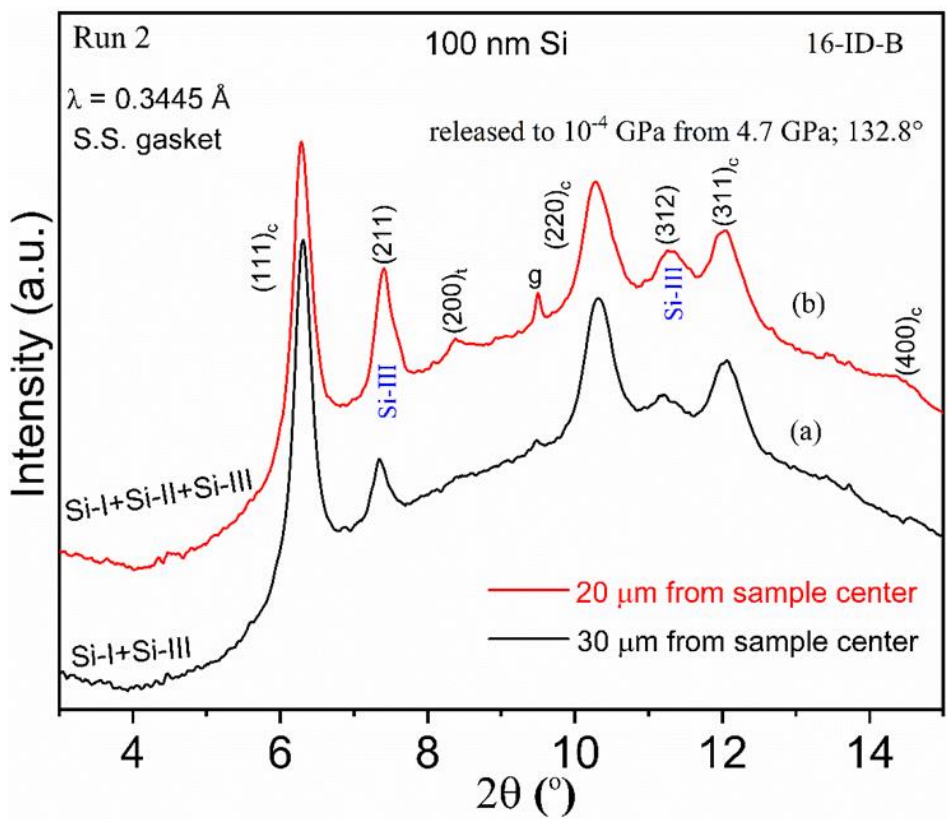

**Fig. S3: XRD patterns of pressure released sample of 100 nm Si after compression and shear inside RDAC.** XRD patterns of pressure released sample compressed and sheared at 4.7 GPa and 132.8° inside RDAC. (a) Si-I+Si-III phase coexistence at r = 30 µm (b) Si-I+Si-II+Si-III phase coexistence at r = 20 µm. 'g' is the gasket peak. The subscripts 'c' and 't' of the Bragg planes indicate cubic diamond and tetragonal phases. The Si-II phase is recovered for the first time in static high pressure experiments.

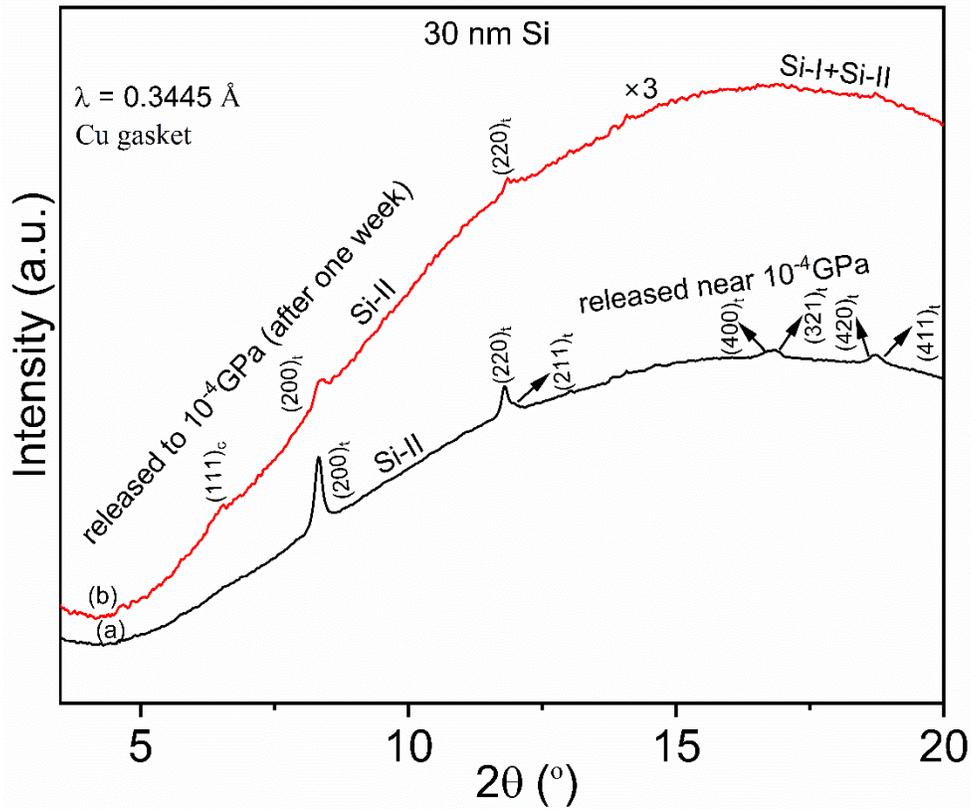

**Fig. S4. XRD patterns of sample of 30 nm Si after non-hydrostatic compression and complete pressure release.** (a) Immediately after pressure release to $10^{-4}$ GPa (ambient) and (b) after one week. The ×3 indicates that the XRD pattern (b) is magnified three times. The subscripts 'c' and 't' of the Bragg planes represent cubic diamond (Si-I) and tetragonal (Si-II) phases. The result represents the first retaining Si-II at ambient pressure.

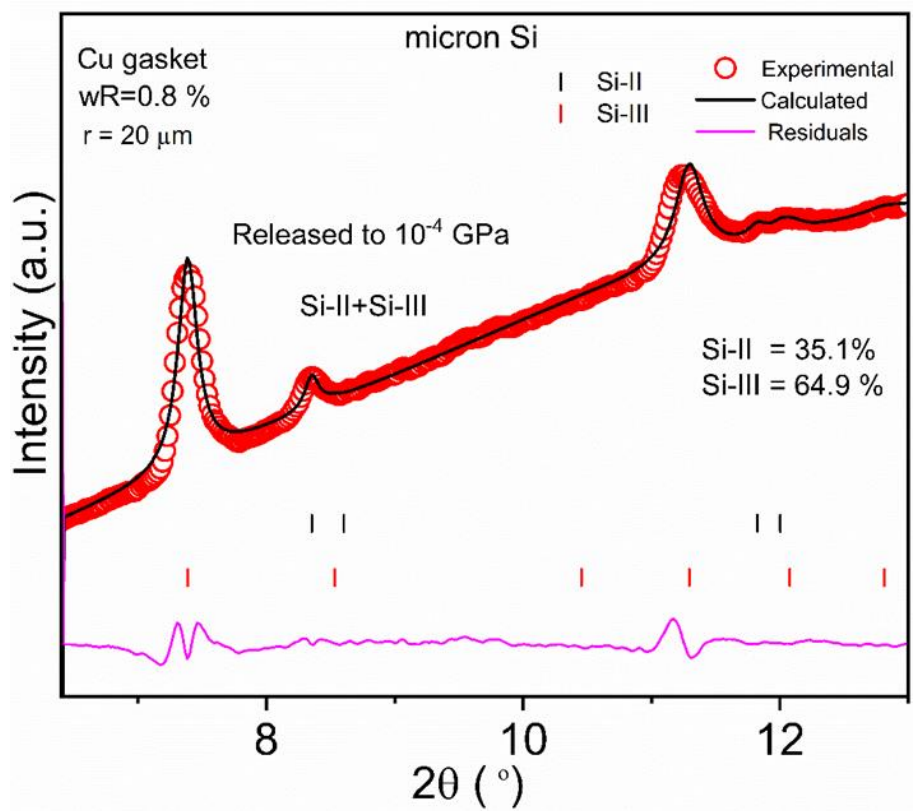

**Fig. S5. Rietveld refinement of pressure released sample of micron-Si.** Rietveld refinement of micron-size silicon at r = 20 µm after pressure release to $10^{-4}$ GPa. The phase fractions (in %) of Si-II and Si-III phases are shown. The Si-II phase is recovered for the first time in static high pressure experiments.

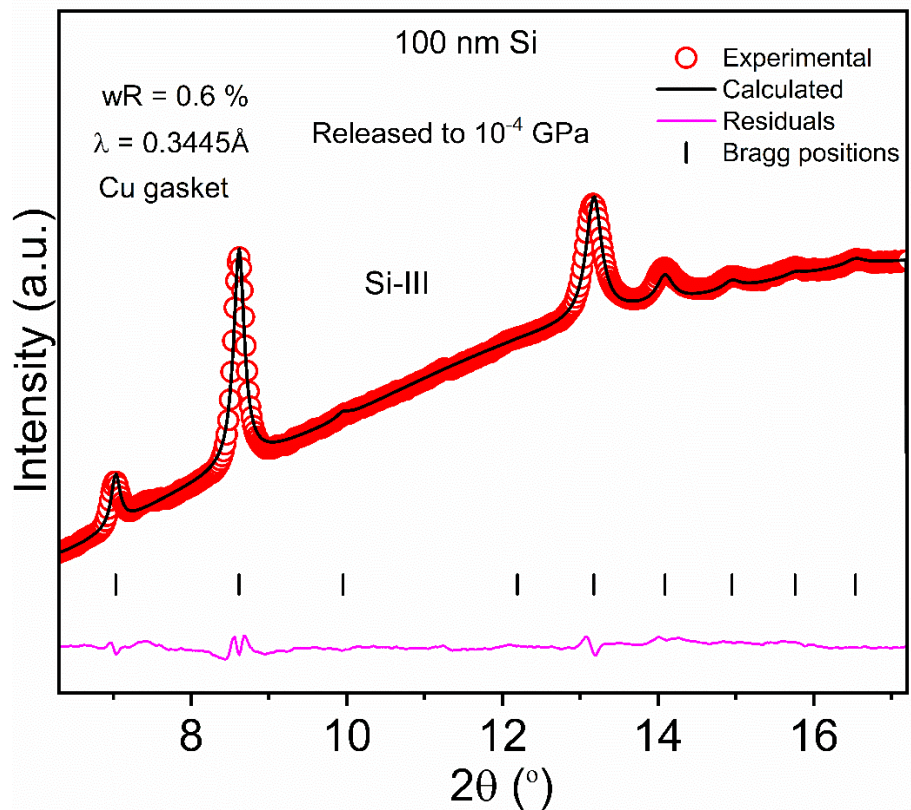

**Fig. S6. Rietveld refinement of pressure released 100 nm Si sample after plastic compression.** The non-hydrostatic compression of 100 nm Si inside DAC above 20 GPa and slowly reduced to $10^{-4}$ GPa. The pressure-released sample is nanostructured single-phase Si-III, which was never obtained before.